\documentclass[10pt,conference]{IEEEtran}

\usepackage{cite}
\usepackage{graphicx}
\usepackage[cmex10]{amsmath}

\begin{document}
%
\title{A Multiobjective Optimization Framework for Routing in Wireless
Ad Hoc Networks}

\author{\IEEEauthorblockN{Katia Jaffr\`es-Runser\IEEEauthorrefmark{1}\IEEEauthorrefmark{2},
Cristina Comaniciu\IEEEauthorrefmark{1} and
Jean-Marie Gorce\IEEEauthorrefmark{2}}
\IEEEauthorblockA{\IEEEauthorrefmark{1}Dept. of Electrical and Computer Engineering,
Stevens Institute of Technology, Hoboken, New-Jersey 07030, USA
\\ Email: \{krunser, ccomanic\}@stevens.edu}
\IEEEauthorblockA{\IEEEauthorrefmark{2}Universit\'e de Lyon, INRIA,
INSA-Lyon, CITI, F-69621, FRANCE\\
Email: {jean-marie.gorce}@insa-lyon.fr}
}


\maketitle

\begin{abstract}
Wireless ad hoc networks are seldom characterized by one single performance metric, yet the current literature lacks a flexible framework to assist in characterizing the design tradeoffs in such networks. In this work, we address this problem by proposing a new modeling framework for routing in ad hoc networks, which used in conjunction with metaheuristic multiobjective search algorithms, will result in a better understanding of network behavior and performance when multiple criteria are relevant. Our approach is to take a holistic view of the network that captures the cross-interactions among interference management techniques implemented at various layers of the protocol stack. The resulting framework is a complex multiobjective optimization problem that can be efficiently solved through existing multiobjective search techniques. In this contribution, we present the Pareto optimal sets for an example sensor network when delay, robustness and energy are considered. The aim of this paper is to present the framework and hence for conciseness purposes, the multiobjective optimization search is not developed herein.
\end{abstract}


\section{Introduction}\label{sec:introduction}
Wireless ad hoc or sensor networks often operate in difficult environments and require several performance criteria to be satisfied, related to timely, reliable, and secure information transfer. To ensure information transfer across a network, one of the key elements is the selected routing protocol whose design poses significant challenges. In such networks, cooperation among all layers of the protocol stack should be enlisted to deal with channel impairments, and thus the design of a routing protocol should be viewed in the context of its interactions with other interference management techniques implemented at other layers of the protocol stack.

To further add to the list of design challenges, it is seldom possible to ``equally optimize" all desirable performance criteria, as some of them may be antagonistic in nature. From a myriad of possible operating points, which one is ``more optimal"? Understanding the tradeoffs involved with respect to various performance metrics will not only lead to a better design, but also will allow for the selection of a set of possible operating points (characterized by various tradeoffs) to enable a graceful degradation of the network performance as the channel conditions worsen.
While significant work has been done for routing in wireless ad hoc or sensor networks, no integrated design framework exists to address the many facets of the problem described above.

There is a significant effort to characterize the theoretical performance of ad hoc wireless networks. Most of it is focused on their theoretical capacity, which has been assessed by several landmark papers under various assumptions \cite{gupta2000, toumpis2003, comaniciu2006, wang2008}. However, none of these works directly supports a practical implementation of a routing algorithm, and they lack a general view of multiple objective tradeoffs - though some of them do consider the impact of the end-to-end delay on capacity.
On the other hand, there is vast literature on designing routing protocols optimized for various specific criteria and specific network instances (e.g. \cite{senouci2004, vassileva2007} and the references within). It is very hard to compare the quality of these solutions as no benchmarks for multiple criteria performance routing exist. Limited work exists on designing multiobjective (MO) routing \cite{kotecha2007}, and again the network scenarios used for optimization are very application specific.

Finally, for networks operating in harsh environments, characterizing cross-layer interactions is essential as for instance in implementing interference mitigation at all layers of the protocol stack. Several seminal works have been written on various aspects of cross-layer design (e.g. \cite{comaniciu2006, chiang2007}).

Understanding the tradeoffs involved with various routing solutions will enable adaptive resource management across layers and nodes, leading to a more accurate "local to global performance mapping" for practical routing protocol design.
Our main contributions in this work are two-fold:
\begin{itemize}
\item Propose a general cross-layer framework network model, capable of capturing the impact and interaction of a wide range of interference and resource management techniques for various channel conditions;
\item Formulate a multiobjective routing optimization problem by defining appropriate evaluation functions for criteria such as: robustness of information transfer, end-to-end delay, and energy consumption.
\end{itemize}

The multiobjective routing optimization problem described in the following can be solved using existing multiobjective search techniques \cite{jaffresChapter08}. However, the description of such a heuristic is out of the scope of this paper and will be addressed in later works.
The paper is organized as follows. In Section \ref{sec:crossLayerFramework} we present our cross-layer framework based on a probabilistic network model. Section \ref{sec:MOproblem} formulates routing in an ad hoc network as a multiobjective optimization problem and Section \ref{sec:SensorNetworks} provides a first formulation applied to sensor networks. Results for a simple problem instance are then given in Section \ref{sec:results} to illustrate our modeling framework and Section \ref{sec:conclusion} concludes the paper.

\section{A Cross-layer Framework for Network Modeling}\label{sec:crossLayerFramework}

\subsection{Probabilistic network model}\label{subsec:ProbabilisticModel}

Our proposed model considers a probabilistic network which is characterized by two probability measures: link and node probability. These two parameters completely characterize the network and capture cross-layer interactions.

{\bf The node probability ($\chi_i$)} captures the availability of node $i$ for routing purposes, i.e. the probability that node $i$ re-broadcasts a received packet. The node probability has two components ($\chi_i = \xi_i \cdot x_i$), one that is determined by the environment and protocol implementations at adjacent layers, $\xi_i$, (e.g. congestion models, node failures, security risks, energy levels), and one component $x_i$ that corresponds to network routing choices, which we aim to optimize in the multiobjective routing framework.

\begin{figure}
\centering
  \includegraphics[width=3.2in]{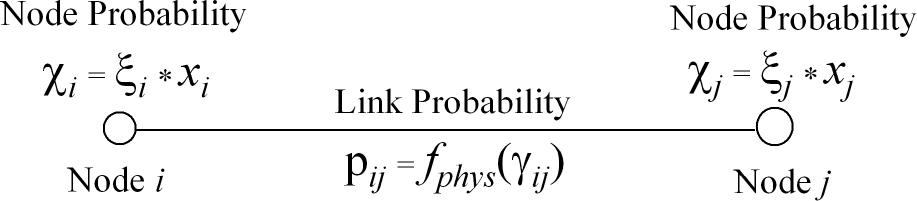}
        \caption{Node and link probabilities on a link $(i,j)$.}
    \label{fig:linkProbNetwork}
\end{figure}

{\bf The link probability ($p_{ij}$)} captures the link availability, i.e., the probability of a successful transmission over a link $(i,j)$. Characterization of the link probability is impacted by impairments and enhancements at various layers of the protocol stack such as fading at the physical layer or congestion at the MAC layer. Both node and link probabilities are illustrated in Fig.~\ref{fig:linkProbNetwork}.

Both node and link probability measures are strongly related due to the nature of the wireless channel. Hence, once the node probabilities $\chi_i$ are set, the activity of every node of the network is fixed and the interference distribution can be completely determined given the nodes's activity on the wireless channel. As a consequence, the link probabilities can be computed as a function of the signal to noise and interference ratio (SINR). Once link and node probabilities are available, various performance metrics such as delay, robustness or energy consumption can be calculated for various transmission schemes (unicast, multicast, broadcast, anycast, etc...).

In the following, we consider the set of node probabilities as the variables of the network optimization problem.
Finding the best possible routing with respect to one particular criterion reduces to the problem of selecting the set of node probabilities that optimizes one particular objective of the network.
Within a multiobjective perspective, solving the network optimization problem requires finding the set of Pareto-optimal solutions that concurrently optimizes several performance metrics of the network.

To illustrate our framework, we consider here a network where the nodes are independent and randomly distributed according to a random point process of density $\rho$ over a disk $\mathcal{D}$.
The communication between any two nodes is performed in a half-duplex mode over a single to multi-hop path.
The bandwidth of the channel is divided into $R$ resources (time slots, frequencies or codes). For clarity purposes, we present this model in the context of time-multiplexing.

This paper concentrates on a single flow but our framework can be extended to multiple flows since the proposed interference model accurately accounts for all the nodes transmitting in the network.
Hence, one source transmits a constant traffic in one of the $R$ time slots.
A relay does not keep track of the packets already transmitted and consequently may forward the same packet several times. However, a node relays the packets in the order they are received in one of its available resources. If several packets are received in the same frame it can only transmit the proportion of packets its global transmission probability $x_i$ allows. The packets that the node can not forward are dropped.
The maximum number of hops $H_M$ a packet can travel in the network is also fixed.

\subsection{Link probabilities}\label{subsec:linkProba}

A realistic link $(i,j)$ in time slot $r$ is characterized by its transmission probability $p_{ij}(r)$, which is a function of the statistical distribution of the SINR at the location of the destination node $j$. Such a computation captures the cross-layer impact of the routing decision on the physical layer performance since the activity of all the nodes of the network are accounted for statistically in the model.
The following are some preliminary definitions and notations that are needed to define the link probability:

\paragraph*{\bf Pathloss attenuation factor}
$a_{ij}$ reflects the attenuation due to propagation effects between node $i$ and $j$. In our simulations, the simple isotropic propagation model is considered.

\paragraph*{\bf Interference}
Since we consider time-multiplexed channels, interference only occurs between transmissions using the same channel at the same time. Hence, the power of interference $I_{ij}(r)$ on a link $(i,j)$ using resource $r$ and computed at node $j$ is defined by:
\begin{equation}
\label{eq:interf}
{\it I}_{ij}(r) = \sum_{k=1}^K P_k~a_{kj}~~{\rm for}~k \neq i
\end{equation}
\noindent where $K$ is the number of interfering signals in resource $r$.

\paragraph*{\bf SINR}
The SINR between any two nodes $i$ and $j$ in resource $r$ is given by:
\begin{equation}
\overline{\gamma}_{ij}(r)=\displaystyle{\frac{P_{ij}}{N_0+I_{ij}(r)}}
\label{eq:sinr}
\end{equation}
where $P_{ij}$ is the power received in $j$, $I_{ij}(r)$ is the interference power on the link and $N_0$ the noise power density.
We have $P_{ij}=P_i~a_{ij}$ for a fixed nominal transmission power $P_i$ and a pathloss attenuation factor $a_{ij}$.

\paragraph*{\bf Packet error rate (PER)}
For a specific value of SINR $\gamma$, the packet error rate $PER$ can be computed according to:
\begin{equation}
PER(\gamma) = 1-\left[1-BER(\gamma)\right]^{N_b} 
\label{eq:PER}
\end{equation}
\noindent where $N_b$ is the number of bits of a data packet and $BER(\gamma)$ is the bit error rate for the specified SINR per bit $\gamma$ which depends on the physical layer technology and the statistics of the channel.
Results are given for an AWGN channel and a BPSK modulation without coding where $BER(\gamma)= Q\left( \sqrt{2\gamma}\right)= 0.5 *{\rm erfc}(\sqrt{\gamma})$.

\paragraph*{\bf Transmission rate}
The activity of a network node in a channel $r \in [1,..,R]$ is given by its transmission rate $\tau_i(r) \in [0,1]$ in that particular channel. This rate is defined as the percentage of time a node $i$ transmits using resource $r$.

\paragraph*{\bf Additional Notations}
A node $i$ is said to be active in the network if $\sum_r \tau_i(r)>0$, and

- $M$ gives the number of active nodes of the network,

- An interfering set on a link $(i,j)$ is a set of $K \leq M-1$ active nodes,

- $\mathcal{L}_{-i}$ refers to the set of all possible interfering sets and has a cardinality of $L=\sum_{k=1}^{M-1}\left(^{M-1}_{~~k}\right)+1$.

\vspace{\baselineskip}

\paragraph*{\bf The link probability} $p_{ij}(r)$ depends on the distribution of the SINR, and consequently on the distribution of the corresponding packet error rates. It is defined by the equation:
\begin{equation}
p_{ij}(r) = \sum_{l=1}^L \left[ 1-PER_l(r)\right].\mathbf{P}_l(r) 
\label{eq:linkProbaPER}
\end{equation}
\noindent where the index $l$ represents one of the $L$ interfering sets. Consequently, $\gamma_l(r)$ is the SINR experienced because of the interfering set $l$ on the link $(i,j)$ for the resource $r$ and $PER_l(r)$ is the corresponding PER. The SINR can be computed according to Eq.~\eqref{eq:sinr} considering the $K$ interfering links of $l$ and the PER according to Eq.~\eqref{eq:PER}.

$\mathbf{P}_l(r)$ is the probability for the link $(i,j)$ to experience the interference distribution $l$ in resource $r$, i.e. the probability that the nodes of the interfering set $l$ are transmitting concurrently and the others are not. Hence, this probability for a link $(i,j)$ is given by:
\begin{equation}
\mathbf{P}_l(r) = \prod_{k=1}^K \tau_k(r)~\cdot~\prod_{m=1}^{M-K-1}(1-\tau_m(r))
\label{eq:probaPER}
\end{equation}
In Eq.~\eqref{eq:probaPER}, $\prod_{k=1}^K \tau_k(r)$ gives the probability that the $K$ active nodes of the interfering set $l$ are transmitting and $\prod_{m=1}^{M-K-1}(1-\tau_m(r))$ the probability that the $M-K-1$ other active nodes are not.

\subsection{Node probabilities and transmission rate}\label{subsec:relationTauX}

The variables of our model are the probability $\chi_i = \xi_i \cdot x_i$ for each node $i$ to re-transmit a received message. In the following, we consider that $\xi_i=1$ to simplify our model. Hence, the main variable is the `forwarding probability' $x_i$.
There is no notion of routing paths herein and a packet sent by a source may use one or more paths in parallel to reach the destination.
For $x_i=1$ each received packet by node $i$ is forwarded. For $x_i<1$ node $i$ drops the packets with probability $1-x_i$. Values of $x_i \in~]1,R]$ are not allowed yet as they imply that node $i$ transmits several copies of the same packet.

As stated earlier, the transmission rate $\tau_i(r)$ in resource $r$ is a function of the node probability $x_i$ but also depends on the amount of traffic coming into node $i$, which is a function of the activity of the other nodes of the network. As a consequence, computing the values of $\tau_i(r)$ knowing the $x_i$ values is intractable since determining the $\tau_i(r)$ requires the knowledge of the link probabilities which are themselves a function of the $\tau_i(r)$ values. However, the reverse approach where the variables $x$ are expressed as a function of the $\tau_i(r)$ can be easily derived as stated below. Hence, such a reverse approach leads to the use of the transmission rates as the variables of our multiobjective optimization problem instead of the forwarding probabilities.
This reverse approach represents an important contribution of our cross-layer model since it captures an exact picture of the interference distribution at the physical layer and determines the corresponding node forwarding probability $x_i$ at the routing level.

\paragraph*{\bf Relationship between $x_i$ and the $\tau_i(r)$}

Given the values of $\tau_i(r), \forall r \in [1..R], i \in [1..N]$, we can define the quantity of information coming from all the neighbors of node $i$ (except from the destination) by:
\begin{equation}
q_i = \sum_{k\neq \{i,D\}}\sum_{r} p_{ki}(r).\tau_k(r).v_{ki}
\end{equation}
where $p_{ki}(r).\tau_k(r).v_{ki}$ is the probability that a packet arrives in node $i$ from node $k$ in resource $r$.

The variable $v_{ki}$ is introduced to represent the usefulness of the link (k,i) with respect to the maximum number of hops constraint. Hence, if no data can arrive from neighbor $k$ because the hop count $h$ for all the packets $k$ received is already equal to $H_M$, we have $v_{ki}=0$. On the contrary, we have $v_{ki}=1$ if $k$ only receives packets with a number of hops $h<H_M$. If $k$ receives packets with both $h<H_M$ and $h=H_M$, $v_{ki}$ represents the proportion of packets being retransmitted.

The quantity of information going out of $i$ is given by the sum of the $\tau_i(r)$ over all the time slots. Hence, we can determine the global forwarding probability of $i$ to be:
\begin{equation}
x_i = \frac{\sum_{r} \tau_i(r)}{ \sum_{k\neq \{i,D\}}\sum_{r} p_{ki}(r).\tau_k(r).v_{ki}}
\label{eq:xi}
\end{equation}

\section{A multiobjective optimization problem}\label{sec:MOproblem}

The performance of most wireless networks can be assessed with regards to various criteria such as throughput or capacity, end-to-end transmission delay, overall energy consumption or transmission robustness.
The purpose of the multiobjective framework presented in this work is to determine, given a network and a communication pattern, what kind of trade-offs arise between chosen performance metrics when varying the routing strategies. It relies on the cross-layer probabilistic network model presented in Section \ref{sec:crossLayerFramework}.

\subsection{Variables of the Multiobjective (MO) Framework}\label{subsec:variables}

The routing strategies are the variables of our multiobjective optimization problem and a solution is defined by:

{\bf Definition 1}
A solution $\mathcal{S}$ of the MO framework is defined by the set of transmission rates $\tau_i(r) \in [0,1]$ used by each node $i$ on each resource $r$:
\begin{equation}
    \mathcal{S}=\left\{\tau_i(r)\right\}_{i \in [1,..,N], r \in [1..R]}
\end{equation}
The set of node probabilities $x_{i, i\in[1..N]}$ is derived according to Eq.\eqref{eq:xi} and represents the routing strategy of the network.
Each variable $\tau_i(r)$ takes its values in a discrete set $\Gamma$ of size $T=|\Gamma|$.
As a consequence, the solution space is derived as:
\begin{equation}
|\mathcal{S}| = \sum_{m=0}^N \left(_{~~m}^{N-2}\right) T^{R.m}
\end{equation}
In order to reduce the size of this very big search space, we only consider solutions where at least one cumulative time slot per node is available in the frame, i.e. $
s.t.~~\forall i\in [1,N], ~\sum_{t=1}^R \tau_i(r)\leq R-1$. The solutions that do not meet this constraint are usually very bad solutions since at least one of the nodes of the solution is transmitting in all its time slots preventing a failure free packet reception.

Using this definition of a routing strategy, a solution may reflect various features: it can be single-hop or multi-hop, single path or multi-path, probabilistic or deterministic.
The aim of our MO framework is to obtain the set of Pareto-optimal routing strategies of the MO problem. A Pareto-optimal set is composed of all the non-dominated solutions of the MO problem with respect to the performance metrics considered. A solution $A$ dominates a solution $B$ for a $n-$objective MO problem if $A$ is at least as good as $B$ for all the objectives and $A$ is strictly better than $B$ for at least one objective.

\section{A first application to Sensor Networks}\label{sec:SensorNetworks}

We propose in the following to assess the performance of a wireless sensor network (WSN) by capturing the trade-offs that arise between end-to-end robustness, overall energy consumption and end-to-end delay. These criteria are prevalent since providing a maximal network throughput is usually not the main task of a WSN.
The criteria are defined for a single source-destination pair $(S,D)$.

\subsection{Robustness criterion}\label{subsec:robustness}
Robustness is defined as the probability that a message emitted at $S$ successfully arrives at $D$ in at most $H_M$ hops. The robustness criterion is given by:
\begin{equation}
f_R=\mathcal{P}(T^{H_M}_{SD})
\label{eq:robustness}
\end{equation}
For any two nodes $i$ and $j$ of the network, $T^H_{ij}$ represents the event that a message transmitted by $i$ successfully arrives in $j$ in at most $H$ hops.
Our aim is to maximize $\mathcal{P}(T^{H_M}_{SD})$.

{\bf Definition 2:} Global link probability.

For a link $(i,j)$, the global link probability $p_{ij}$ is the probability that a message arrives with success at node $j$. It is given by:
\begin{equation}
p_{ij} = \sum_{r=1; \tau_i(r)\neq 0}^R p_{ij}(r)~\frac{\tau_i(r)}{\sum_{r} \tau_i(r)}
\label{eq:linkProba}
\end{equation}
\noindent where $p_{ij}(r)$ is the link probability between $i$ and $j$ for resource $r$ ({\it cf}. Eq.~\eqref{eq:linkProbaPER}) and $\tau_i(r)/\sum_{r} \tau_i(r)$ the probability for the packet to be sent using $r$.

{\bf Definition 3:} Robustness probability.

$\mathcal{P}(T^{H_M}_{SD})$ is the probability that the message arrives successfully in $D$ in at most $H_M$ hops and is given by:
\begin{equation}
\mathcal{P}(T^{H_M}_{SD}) = 1- \prod_{h=1}^{H_M} (1-\mathcal{P}(T_{SD}|H=h))
\label{eq:probaRobustness}
\end{equation}
\noindent where $\mathcal{P}(T_{SD}|H=h)$ is the probability for a packet to arrive in $h$ hops at $D$. For $h=1$, $\mathcal{P}(T_{SD}|H=1) = p_{SD}$, the successful transmission probability on the link $(S,D)$ following Eq.~(\ref{eq:linkProba}).
For $h>1$, we have:
\begin{equation}
\mathcal{P}(T_{SD}|H=h) = 1 - \prod_{j=1}^{N_S} \left[1 - p_{Sj}~x_j~\mathcal{P}(T_{jD}|H=h-1)\right]
\label{eq:probaRecursif}
\end{equation}
\noindent with $N_S$ the number of possible first hop relays of $S$; $p_{Sj}$ the link probability between $S$ and its neighbor $j$; $\mathcal{P}(T_{jD}|H=h-1)$ the probability to reach $D$ in $(h-1)$ hops and $x_j$ the forwarding probability of $j$.
The set of $N_S$ relays is given by all the nodes different from $S$ that are active in at least one of the time slots in the current solution (i.e. having $\sum_{t=1}^R (x_i^t)>0, ~i \neq \{j, S\}$).

To reduce the computation complexity of the robustness probability, a restricted set $N_S$ of first hop relays may be considered but the loss in terms of accuracy is hard to quantify. Therefore, we rather introduce a {\it link threshold value $\mathcal{P}_{th}$} computed for each path made of $h$ hops. While recursively calculating $\mathcal{P}(T_{SD}|H=h)$, if the probability of a path gets lower than $\mathcal{P}_{th}$, the recursion is stopped for that particular path and its contribution to $\mathcal{P}(T_{SD}|H=h)$ is set to zero.

\subsection{Delay criterion}\label{subsec:delay}
The end-to-end delay is the sum of the times spent at each relay on a multi-hop path where each relay introduces a delay of 1. The criterion $f_D$ is defined by:
\begin{equation}
f_D = R\cdot\displaystyle\sqrt{ \sum_{h=1}^{H_M} (h-1)^2 .R_h }
\label{eq:delay}
\end{equation}
The quantity $(h-1)$ is the delay needed by a packet to arrive in $h$ hops using $(h-1)$ relay nodes. The scaling factor $R$ represents the delay induced by the $R$ resources. $R_h$ is the probability that the packet arrived in exactly $h$ hops and did not arrive in 1, or 2… or $(h-1)$ hops. For $h=1$, we have $R_h=P(T_{SD}|h=1)$ and for $h>1$:
\begin{equation}
R_h = \mathcal{P}(T_{SD}|H=h).\prod_{i=1}^{h-1}(1-\mathcal{P}(T_{SD}|H=i))
\end{equation}
If no route exists between $S$ and $D$ then $f_D=+\infty$.

\begin{figure}
\begin{center}$
\begin{tabular}{|c|c||c|c|}
\hline
    \footnotesize{Transmission Power}    &  \footnotesize{151mW}       & \footnotesize{$N_0$}         & \footnotesize{-154dBm/Hz} \\ \hline
    \footnotesize{Bandwidth}     & \footnotesize{1Mbps}        & \footnotesize{$f$} & \footnotesize{$2.4GHz$}   \\ \hline
    \footnotesize{Pathloss exponent $\alpha$}        & \footnotesize{3} & \footnotesize{Channel Model} & \footnotesize{AWGN}\\ \hline
        \footnotesize{Antenna gains}&        \footnotesize{$G_T$=$G_R$=1} & \footnotesize{Modulation} & \footnotesize{BPSK}\\\hline

\end{tabular}$
\end{center}
\caption{Propagation and physical layer parameter values.}
\label{tab:values}
\end{figure}

\subsection{Energy criterion}\label{subsec:energy}
The energy criterion $f_E$ is given by the total {\it forwarding energy} needed for a packet sent by $S$ to reach $D$.
We do not account for the energy spent by the initial transmission in $S$. The reception (resp. transmission) of a packet at node $j$ in resource $r$ consumes $e_j^R(r)$ (resp. $e_j^T(r)$).
Hence, the energy criterion is defined as:
\begin{equation}
f_E = \sum_{h=1}^{H_{M}} \mathcal{E}(T_{SD}|H=h)
\label{eq:energy}
\end{equation}
\noindent where $\mathcal{E}(T_{SD}|H=h)$ is the total energy needed by the $h$-hop communications between $S$ and $D$ defined by:
\begin{multline}
\label{eq:energyRecursif}
\mathcal{E}(T_{SD}|H=h) =
        \\ \sum_{j=1}^{N_S}
                \left(
                    p_{Sj}.e_j^R +
                   p_{Sj}.x_j.\left[
                        e_j^T + \mathcal{E}(T_{jD}|H=h-1)\right]
                \right)
\end{multline}
In Eq.~\eqref{eq:energyRecursif}, $p_{Sj}.e_j^R$ is the energy consumed for a packet reception by the neighbor $j$ of $S$; $p_{Sj}.x_j.e_j^T$ is the energy consumed for the packet transmitted by neighbor $j$ and $p_{Sj}.x_j.\mathcal{E}(T_{jD}|H=h-1)$ is the total energy consumed by the following possible paths made of $(h-1)$ hops between neighbor $j$ and the destination.
For $h=1$, $\mathcal{E}(T_{SD}|H=1)=0$ since the energy in $S$ is not accounted for.

\section{First Results}\label{sec:results}
\begin{figure*}
\centering
  \includegraphics[width=6.5in]{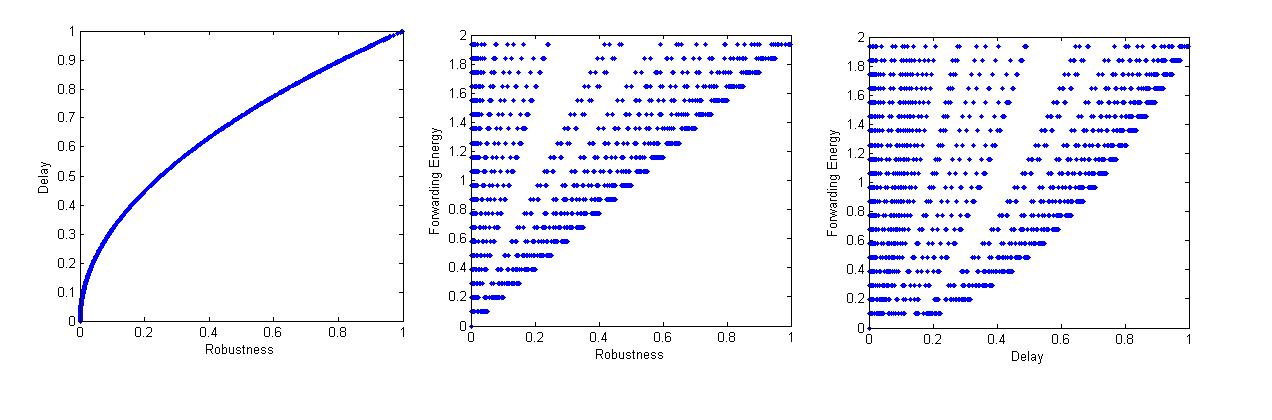}
        \caption{Representation of the projections of the Pareto-optimal set for the 1-relay problem.
        }
    \label{fig:ParetoProbaM1}
\end{figure*}

\subsection{$M$-Relay problem}
The results presented in this section are obtained for a small problem instance for two reasons. First, we are able to determine the complete Pareto-optimal set of solutions using an exhaustive search. Secondly, such a problem can be easily analyzed and provides a first illustration of our multiobjective framework. We will tackle bigger instances using multiobjective optimization algorithms \cite{jaffresChapter08}.

In the following, the network is composed of $N=333$ nodes uniformly distributed with density $\rho=0.004$ over a disk $\mathcal{D}$ of radius $R_{\mathcal{D}}$.
The distance between $S$ and $D$ is of about 215 meters. To reduce border effects, $S$ and $D$ are selected within a radius $R_{\mathcal{C}}<<R_{\mathcal{D}}$ which ensures that the power of a node at distance $R_{\mathcal{C}}$ is below the noise power for the nodes located at distance $R_{\mathcal{D}}$.
We consider $R=2$ time slots and use a probabilistic discrete variable space where $\tau_i(r)$ takes its values in the set $\Gamma=\{0, 0.05, 0.1, \dots 0.9, 0.95, 1.0\}$ of $|\Gamma|=21$ elements. A link robustness threshold of $\mathcal{P}_{th}=10^{-10}$ is set.
Propagation and physical layer parameters are summarized in Fig.~\ref{tab:values}.

The dimension of the search space can be modified by setting a maximum number of forwarding nodes $M$ in a solution $\mathcal{S}$. This sub-problem is addressed in the following as the $M$-relay problem instance.

\subsection{Pareto-optimal set for the 1-relay problem}
In this problem instance, we set $M=1$ and $H_M=2$. In that particular case, the search space has a dimension of 72813 solutions and the Pareto-optimal set is obtained with an exhaustive search.

For this instance, the direct link $(S,D)$ is very weak. A robustness of only $\mathcal{P}(T^{H_M}_{SD})=0.0003$ is achieved with a delay of $f_D=0$ and an energy of $f_E=0$.
Only 24820 solutions fulfill the constraint $x_i\leq1$ that forbids a node to duplicate packets.
Among these solutions, 3855 solutions are Pareto-optimal, representing respectively about 5\% and 15\% of the whole and the constrained solution space.
For all the Pareto-optimal solutions the relay never transmits in the first time slot concurrently with the source. Hence, the model suggests that the Pareto-optimal set for this case is composed of solutions that minimize interference.

The performance of the Pareto-optimal set of solutions is represented in Fig.~\ref{fig:ParetoProbaM1} in the space defined by the three evaluation functions. For clarity purposes, the projections of the Pareto set on the robustness/delay, robustness/energy and the delay/energy planes are displayed.
The plots of Fig.~\ref{fig:ParetoProbaM1} show that an improved robustness is obtained at the price of an increase in delay and energy. The trade-off between robustness and delay can be easily understood since higher robustness is achieved when the relay contributes with a higher forwarding probability $x_i$, inducing an increase in delay. Similarly, an increase of $x_i$ triggers an accrued average energy consumption since the relay is forwarding packets more often.

The energy consumption for all the Pareto-optimal solutions belongs to a discrete set of 21 energy levels which is a direct consequence of the 21 values of $\tau_i(r)$ defined in this problem instance. Hence, the definition of a continuous transmission rate variable $\tau_i(r)$ would provide the most precise description of the Pareto set. However, tackling the continuous formulation of our problem is much more challenging and for our study, we will stick to the simpler discrete formulation which still provides a fair representation of the Pareto set.

\begin{figure}
\centering
  \includegraphics[width=2.7in]{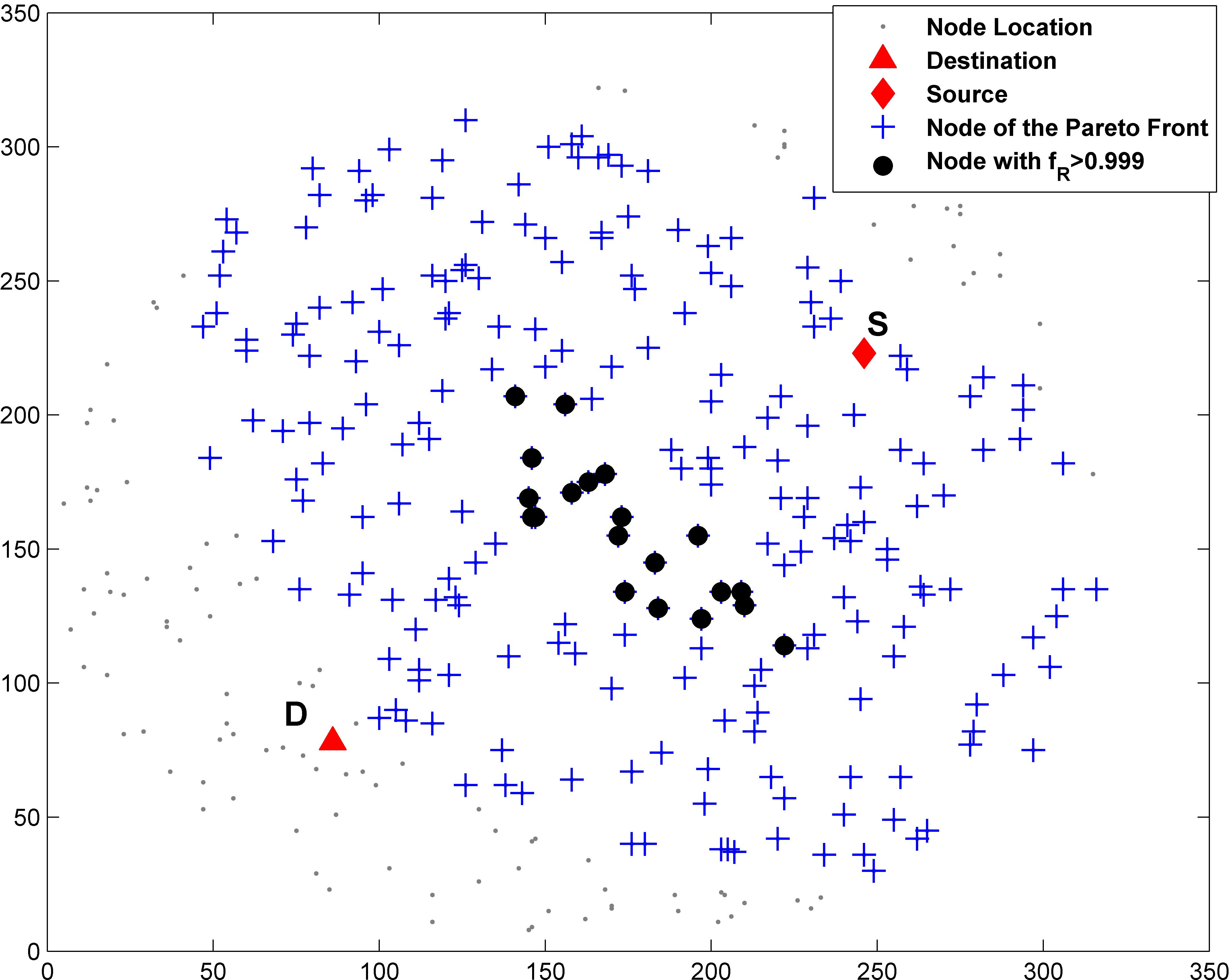}
        \caption{Location of the nodes that provide Pareto-optimal solutions in the network (blue crosses) and of the nodes that provide a near quasi-perfect robustness (full black dots), i.e. $f_R>0.999$.
        }
    \label{fig:allSols}
\end{figure}

The Pareto set is composed of solutions where relays belong to a set of 226 nodes, which represents about two thirds of the number of nodes of the network. The location in the network of these 226 nodes is presented in Fig.~\ref{fig:allSols}. We also highlighted on this figure the relays that provide a near perfect transmission. We can conclude that the relays located in an ellipse near the middle of the $(S,D)$ distance provide the best robustness at the price of the highest delay and energy. The other relays present in the Pareto set provide various trade-offs depending on their values of $\tau_i(r)$.

This first simple study shows that the proposed multiobjective probabilistic network model provides a coherent and complete view of the trade-offs that arise between robustness, delay, and energy in our network. A more extensive analysis of the performance of the model has to be performed next by considering a complete solution space and various network topologies. For such instances, the size of the search space prohibits the use of an exhaustive search of the Pareto set. Hence, we will concentrate on implementing a combinatorial multiobjective optimization algorithm to obtain the best possible representation of the Pareto-optimal set.

%
%
%

\section{Conclusion}\label{sec:conclusion}
In this paper, we have proposed a novel multiobjective optimization framework for network routing in wireless ad hoc networks.
Our proposed framework consists of a general probabilistic network model capable of capturing the impact and interaction of a wide range of resource/interference management techniques, and various channel conditions and network scenarios. Used in conjunction with metaheuristic optimization techniques, this framework provides an efficient tool to capture the trade-offs between different performance metrics and obtain bounds on the achievable performance of routing for a single source-destination transmission.
Preliminary results were obtained in characterizing the delay, robustness, and energy tradeoffs for a 2-hop sensor network model.
Future work will extend the model to consider more complex network scenarios, such as to account for various network topologies, to consider multiple concurrent flows in the network, and to use more refined cross-layer interactions and interference models.

\section*{Acknowledgments}
This work was supported in part by the Marie Curie OIF Action of the European Community's Sixth Framework Program (DistMO4WNet project) and by the ONR grant \#N00014-06-1-0063. This article only reflects the author's views and neither the Community nor the ONR are liable for any use that may be made of the information contained herein.

\end{document}